\documentclass[aps,pre,reprint,groupedaddress,showpacs]{revtex4-1}
\usepackage{amsmath,amssymb,amsxtra,amsfonts,graphicx,wrapfig}
\usepackage{xcolor}
\usepackage{subfigure}
\usepackage[font=small,labelfont=bf]{caption}
\usepackage{color}
\begin{document}
\title{Reservoir crowding in a totally asymmetric simple exclusion process with Langmuir Kinetics}
\author{Bipasha Pal}
\author{Arvind Kumar Gupta}
\email[]{akgupta@iitrpr.ac.in}
\affiliation{Department of Mathematics, Indian Institute of Technology Ropar, Rupnagar-140001, Punjab, India.}

\date{\today}
\begin{abstract}
We study a totally asymmetric simple exclusion process equipped with Langmuir kinetics with boundaries connected to a common reservoir. The total number of particles in the system is conserved and controlled by filling factor $\mu$. Additionally, crowding of reservoir is taken into account which regulates the entry and exit of particles from both boundary as well as bulk.  In the framework of mean-field approximation, we express the density profiles  in terms of Lambert-W functions and obtain phase diagrams in $\alpha-\beta$ parameter space. 
 Further, we elucidate the variation of phase diagram with respect to filling factor and Langmuir kinetics. In particular, the topology of the phase diagram is found to change in the vicinity of $\mu=1$. Moreover, the interplay between reservoir crowding and Langmuir kinetics develops a novel feature in the form of back-and-forth transition. The theoretical phase boundaries and density profiles are validated through extensive Monte Carlo simulations.

\end{abstract}
\maketitle
\section{Introduction}

Totally asymmetric simple exclusion process (TASEP) is a paradigmatic model to analyze the system of self-driven particles that evolves into non-equilibrium steady states \cite{blythe2007nonequilibrium,macdonald1968kinetics,zia2011modeling}. While the behavior of such systems is usually quite complex, TASEP models are simple enough to be analyzed in great detail. These models were initially proposed to study the kinetics of biopolymerization on nucleic acid forms \cite{macdonald1968kinetics}. The model includes unidirectional particle jumps from one site to the next site on a one-dimensional lattice while respecting hardcore exclusion principle. Since its inception, TASEP has undergone substantial modifications that mimic various non-equilibrium processes including traffic flow and biological transport \cite{lazarescu2011exact,parmeggiani2003phase,kolomeisky1998phase}.

Many versions of the TASEP have been thoroughly investigated over the years \cite{kolomeisky1998phase,evans2003shock,parmeggiani2004totally,dhiman2014two,hilhorst2012multi,muhuri2011bidirectional,sharma2017phase,dong2007inhomogeneous,dong2008understanding,turci2013transport,schmidt2015defect,shaw2003totally,pierobon2006bottleneck}. These models incorporate Langmuir Kinetics (LK) \cite{popkov2003localization,evans2003shock,parmeggiani2004totally,dhiman2018steady,botto2018dynamical}, multiple lanes \cite{dhiman2014two,hilhorst2012multi,wang2017dynamics,gupta2016collective}, bidirectional movement \cite{muhuri2011bidirectional,sharma2017phase,jelic2012bottleneck,zia2011modeling}, etc., to name a few. The additional dynamics of LK in TASEP pertains to the association/dissociation of particles from the bulk. This inclusion mimics various physical systems and displays remarkable features \cite{parmeggiani2003phase,parmeggiani2004totally,popkov2003localization,evans2003shock,dhiman2018steady,botto2018dynamical}. 
In recent years, an interesting variant of TASEP has evolved which requires the total number of particles in the system to remain conserved. Such a modification effectively reflects the limited resources in biological or physical processes namely protein synthesis, movement of motor proteins, vehicular traffic, etc. \cite{cook2009feedback,cook2013interplay,brackley2012multiple}. In this direction, \textit{Car Garage model} was studied wherein the entry rate is unaffected provided there is atleast one particle in the system \cite{ha2002macroscopic}. Later,  an open TASEP  was incorporated with a global constraint on available number of particles where the entry rate is controlled by the number of particles in the reservoir \cite{adams2008far}. In another variant, the effect of coupling several  open TASEPs to a finite reservoir was discussed \cite{cook2009competition}. These models primarily focus on the effect of finite resources on the rate by which the particles enter the lattice, whether it be boundary or bulk. However, exit rate of particles from the TASEP also can be affected. For instance, consider the vehicles that intend to leave the road and enter into a parking area. It is expected that the vehicles will leave the road at a higher rate when parking area is relatively empty. In a similar manner, one can expect that a crowded parking area would obstruct the path of vehicles and reduce the rate. Consequently, the rate at which the vehicle leaves the road is significantly affected by the occupancy of the parking area. Thus, incorporating the effect of number of particles in the reservoir, termed as \textit{reservoir crowding}, on both entry and exit rates is justifiable. A variant of TASEP that takes reservoir crowding into account thereby affecting  both the entry and exit rate has been studied \cite{haldar2020asymmetric}. This feature has been argued as a mechanism by which the particles avoid the crowding of reservoir.

In this paper, we study a model comprising of a TASEP with LK provided the number of particles in the system remains conserved. Additionally, we consider both the ends of the TASEP to be connected to a reservoir that features reservoir crowding. This enables the particles to be more likely to leave the reservoir and inhibits the tendency to rejoin it \cite{haldar2020asymmetric}.  Our model corresponds to a modified version of Ref.\cite{evans2003shock,parmeggiani2004totally} where  all the rates (except forward hopping with unit rate in bulk) are affected by particle number conservation (PNC). One can also interpret our model as a generalization of Ref.\cite{haldar2020asymmetric}, wherein we allow the association and dissociation of particles in the bulk with rates regulated by number of particles in the reservoir. Our focus is on exploring the consequences of interplay between LK and filling factor in presence of reservoir crowding. In this direction, we intend to employ mean-field arguments to theoretically analyze the system properties including density profiles and phase diagrams. Also, we investigate how the phase diagram is affected when the number of particles in the system are varied. Additionally, we scrutinize the effect of LK rates on the structure of phase diagram. We proceed with the aforementioned motives and study a few more features of the model.
\section{\label{sec:level1}Model}
 We consider a  lattice, denoted by $T$, with total $L$ sites labelled as $i\in\{1,2,\dots,L\}$. The sites $ i=1,L$ define the boundaries of $T$, while the remaining $L-2$ sites are collectively referred to as bulk. Both the ends of $T$ are connected to a  reservoir $R$  having no internal dynamics. The particles from $R$ enter $T$ at entry site $i=1$ with an innate entry rate $\alpha$, hop unidirectionally along $T$ with unit rate following the hardcore exclusion principle. When the particle reaches the exit site $i=L$, it leaves $T$ and rejoins $R$ with an innate exit rate $\beta$.

\begin{figure}[h]
\centering
\includegraphics[height=1.1in]{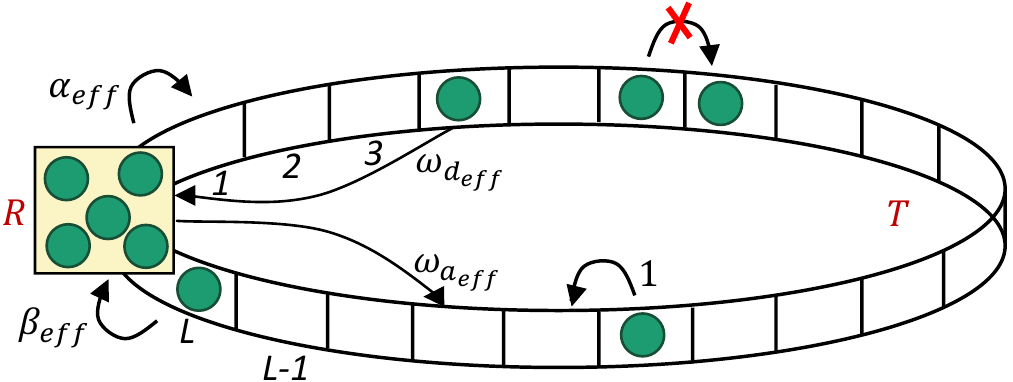}
  \caption{Schematic diagram of the model consisting of lattice $T$ and reservoir $R$. The effective entry and exit rates at the site $1$ and $L$ are given by $\alpha_{eff}$ and $\beta_{eff}$, respectively; $\omega_{a_{eff}}$ and $\omega_{d_{eff}}$ denote the effective attachment and detachment rates in the bulk.}
  \label{fig:model1}
\end{figure}

 Additionally, we consider the innate attachment (detachment)  of particles from $R$ (bulk)  to bulk ($R$) at rate $\omega_a$  ($\omega_d$).  This system can be viewed as a closed TASEP with LK having a special site $L+1$ which corresponds to the reservoir  and violates the exclusion principle (see Fig.\ref{fig:model1}). The TASEP has both conserving and non-conserving dynamics, but the total number of particles in the system, $N_{tot}$, remains conserved owing to the fact that the particles leaving the TASEP from the bulk rejoins the reservoir.

 In our model, the reservoir can be treated as a point reservoir which is connected to both ends of the lattice $T$. The entry and exit rates of particles depend on the number of particles in the reservoir. The availability of particles to enter $T$  affects the entry rate and attachment rate. Correspondingly, the exit rate and detachment rate are influenced by the hindrances occurring due to the particles in reservoir which is termed as crowding effect \cite{haldar2020asymmetric}. Thus, all the rates except the hopping rate in the bulk are dynamically controlled by the number of particles in the reservoir, given by
\begin{equation}
\label{eq:effectiverates}
\begin{aligned}
\alpha_{eff}&=\alpha f(N_r),\\
\beta_{eff}&= \beta g(N_r),\\
\omega_{a_{eff}}&=\omega_a f(N_r),\\
\omega_{d_{eff}}&=\omega_d g(N_r),
\end{aligned}
\end{equation}
where $N_r$ is the instantaneous number of particles in $R$. The functions $f(.)$ and $g(.)$ regulate the entry and exit, respectively, of particles on $T$ and the choice of $f$ and $g$ controls the system dynamics. In a general situation,  larger number of particles in $R$ not only leads to a higher influx of particles in $T$, but also prevents the outflow of particles from $T$ to $R$; reverse is true for the case where $R$ contains less particles. This can be mimicked by considering $f(.)$ and $g(.)$ to be monotonically increasing and decreasing function of $N_r$, respectively, defined as $f(N_r)=N_r/N_{tot}$ and $g(N_r)=1-N_r/N_{tot}$ \cite{haldar2020asymmetric}. These functions are bounded by $0$ and $1$ which results in the effective rates to be bounded by $0$ and the innate rates.

To explore the effect of total number of particles on the system dynamics, we define filling factor $\mu = N_{tot}/L$ \cite{haldar2020asymmetric}. 
For the limiting case $\mu\to\infty$, our model is not analogous to an open TASEP with LK \cite{parmeggiani2004totally} where the rates remain unaffected by the instantaneous number of particles in reservoir. On contrary, in the present model as $\mu\to\infty$,  the effective exit and detachment rates approach to $0$, i.e., $\beta_{eff}\to 0$ and $\omega_{d_{eff}}\to 0$, due to overcrowding. Meanwhile, the effective entry rate and attachment rate converge to the corresponding innate rates, i.e., $\alpha_{eff}\to\alpha$ and $\omega_{a_{eff}}\to\omega_a$.  Thus, at $\mu\to\infty$, our model converges to a special case of an open TASEP with LK where exit rate and detachment rate are $0$ \cite{parmeggiani2004totally}.

\section{\label{sec:level2}Master Equations and Mean-field analysis}
We define $\tau_i$ to be the occupation number of the $i^{th}$ site of the lattice. As the particles obey the hardcore exclusion principle, $\tau_i$ takes only binary values, $0$  or $1$,  depending upon whether the site is vacant or occupied, respectively. The master equation of evolution of density in the bulk is given by
\begin{equation}
\label{eq:master}
\begin{aligned}
\dfrac{d\langle \tau_i \rangle}{dt}=&\langle\tau_{i-1} (1-\tau_i)\rangle+\omega_{a_{eff}}\langle 1-\tau_i\rangle\\&-\langle\tau_i(1-\tau_{i+1})\rangle-\omega_{d_{eff}}\langle \tau_i\rangle,
\end{aligned}
\end{equation}
whereas, at the boundaries, the density evolves according to the following equations
\begin{equation}
\label{eq:leftboundary}
\dfrac{d\langle \tau_1 \rangle}{dt}=\alpha_{eff}\langle 1-\tau_1\rangle-\langle\tau_1(1-\tau_2)\rangle,
\end{equation}
\begin{equation}
\label{eq:rightboundary}
\dfrac{d\langle \tau_L \rangle}{dt}=\langle \tau_{L-1}(1-\tau_L)\rangle-\beta_{eff}\langle\tau_L\rangle,
\end{equation}
where $\langle\dots\rangle$ denotes the statistical average. 
The above equations reduce to those in Ref.\cite{parmeggiani2004totally} for constant functions $f=g=1$ in Eq.\eqref{eq:effectiverates} ie., the effect of reservoir crowding is removed.
Eqs.\eqref{eq:master}, \eqref{eq:leftboundary} and \eqref{eq:rightboundary} cannot be solved in the present form due to the presence of two-point correlators. We employ mean-field approximation which  neglects all types of correlations i.e., $\langle\tau_i\tau_j\rangle=\langle\tau_i\rangle\langle\tau_j\rangle$ and is found to be exact for simple TASEP \cite{kolomeisky1998phase}. 
After defining the average density at site $i$ as $\rho_i=\langle\tau_i\rangle$, we coarse-grain $T$ by introducing a quasi-continuous variable $x=i/L\in[0,1]$  using lattice constant $\epsilon=1/L$  and re-scaled time $t'=t/L$, in the thermodynamic limit. When $\omega_a$ and $\omega_d$ are considered to the independent of $L$, the effect of attachment and detachment is negligible for large but finite systems  $(L\gg1)$. To observe the competition between bulk and boundary dynamics in large systems, the kinetic rates $\omega_a$ and $\omega_d$ must decrease as $L$ increases such that the reduced rates $\Omega_a$ and $\Omega_d$ remain constant with $L$ \cite{parmeggiani2004totally}. Therefore, we  set
\begin{equation}
\Omega_a=\omega_aL,~\Omega_d=\omega_dL.
\end{equation}
On expanding the average density $\rho(x)$ in powers of $\epsilon$ and retaining the terms up to second order, we obtain
\begin{equation}
\label{eq:continuumeq}
\dfrac{\partial\rho}{\partial t}+\dfrac{\partial}{\partial x}\bigg(-\dfrac{\epsilon}{2}\dfrac{\partial\rho}{\partial x}+\rho(1-\rho)\bigg)=\Omega_{a_{eff}}(1-\rho)-\Omega_{d_{eff}}\rho.
\end{equation}
We define the density of reservoir to be $\rho_r=N_r/L$.
In stationary state, Eq.\eqref{eq:continuumeq} yields
\begin{equation}
\label{eq:steadystateeq}
\dfrac{\epsilon}{2}\dfrac{\partial^2\rho}{\partial x^2}+(2\rho-1)\dfrac{\partial\rho}{\partial x}+\Omega_{a}(1-\rho)\frac{\rho_r}{\mu}-\Omega_{d}\rho\bigg(1-\frac{\rho_r}{\mu}\bigg)=0.
\end{equation}
Eq.\eqref{eq:leftboundary} and \eqref{eq:rightboundary} turn into boundary conditions,
\begin{equation}
\label{eq:bc}
\rho(0)=\alpha\frac{ \rho_r}{\mu} \text{ and } \rho(1)=1-\beta\Big(1-\frac{\rho_r}{\mu}\Big),
\end{equation}
 respectively. In the limit $\epsilon\to0$, Eq.\eqref{eq:steadystateeq} yields
\begin{equation}
\label{eq:fode}
\dfrac{\partial J}{\partial x}=\Omega_{a}(1-\rho)\frac{\rho_r}{\mu}-\Omega_{d}\rho\bigg(1-\frac{\rho_r}{\mu}\bigg).
\end{equation}
where $J$ denotes steady-state bulk current and is given by $J=\rho(1-\rho)$. Eq.\eqref{eq:fode} is  a first order differential equation with two boundary conditions, thereby making the problem over-determined. Nevertheless, the solutions of Eq.\eqref{eq:fode} can be defined by utilizing only one of the boundary conditions. We denote the solution satisfying $\rho(0)=\alpha\frac{ \rho_r}{\mu}$ by $\rho_{\alpha}(x)$ and $\rho(1)=1-\beta\big(1-\frac{\rho_r}{\mu}\big)$ by $\rho_{\beta}(x)$.  

Now, to find the value of $\rho_r$, we write the following master equation which regulates the density of reservoir
\begin{equation}
\label{eq:res}
\begin{aligned}
\dfrac{d\rho_r}{dt}=&\beta_{eff} \langle \tau_{L}\rangle+\omega_{d_{eff}}\big(\langle\tau_{2}\rangle+\langle\tau_{3}\rangle+\dots+\langle\tau_{L-1}\rangle\big)\\&-\alpha_{eff}\langle1-\tau_1\rangle-\omega_{a_{eff}}\big(\langle1-\tau_{2}\rangle+\langle1-\tau_{3}\rangle\\&+\dots+\langle1-\tau_{L-1}\rangle\big).
\end{aligned}
\end{equation}
To reduce the parameter space, we consider the innate attachment and detachment rates to be equal i.e., $\Omega_a=\Omega_d=\Omega$. Despite such an assumption, the effective attachment and detachment rates may or may not be equal. After coarse-graining and applying the thermodynamic limit, we write the steady-state equivalent of  Eq.\eqref{eq:res}, and obtain
\begin{equation}
\label{eq:reseq}
\rho(1)\beta\Big(1-\frac{\rho_r}{\mu}\Big)-\frac{\alpha\rho_r}{\mu}\big(1-\rho(0)\big)=\Omega\Big(\rho_r-\mu+\frac{\rho_r}{\mu}\Big).
\end{equation}
When  $\alpha\neq\beta$, the above equation is quadratic whose solutions are given by $\rho_r=A\pm B$
where
\begin{equation}
\begin{small}
\begin{aligned}
A&=\dfrac{\mu\big(\alpha+\beta-2\beta^2+\Omega(\mu+1)\big)}{2(\alpha^2-\beta^2)},\\ B&=\dfrac{\mu\sqrt{(-\alpha-\beta+2\beta^2-\Omega(\mu+1))^2-4(\alpha^2-\beta^2)(\beta-\beta^2+\Omega\mu)}}{2(\alpha^2-\beta^2)}.
\end{aligned}
\end{small}
\end{equation}
From the two different values of $\rho_r$, we use the feasible value depending upon the parameters $\alpha, \beta, \Omega$, and $\mu$.
 For $\alpha=\beta$, Eq.\eqref{eq:reseq} reduces to a linear equation which yields the following reservoir density
\begin{equation}
\label{eq:onerhor}
\rho_r=\dfrac{\mu(-\beta^2+\beta+\Omega\mu)}{-2\beta^2+2\beta+\Omega(1+\mu)}.
\end{equation}

For further analysis, we adopt a methodology that involves transforming  Eq.\eqref{eq:fode} to a form whose solution is already known \cite{parmeggiani2004totally,corless1996lambertw}. Towards this end, we introduce a re-scaled density of the form
\begin{equation}
\label{eq:transform}
\sigma(x)=\dfrac{2\rho-1}{1-\frac{2}{\mu}\int_0^1\rho(x)dx}-1.
\end{equation}
Here $\sigma=0$ corresponds to Langmuir isotherm which depends on filling factor and is given by $\rho_l=\mu/(1+\mu)$. 
By defining the effective binding constant $K_{eff}=\Omega_{a_{eff}}/\Omega_{a_{eff}}$, the expression for $\rho_l$ becomes $K_{eff}/(1+K_{eff})$ which is similar to that in Ref.\cite{parmeggiani2004totally}.
From PNC, we get $N_{tot}=L\int_0^1\rho dx+N_r$ which further yields $\int_0^1\rho dx=\mu-\rho_r$.  It is evident from Eq.\eqref{eq:transform} that $\sigma(x)$ is not defined at  $\rho_r=\mu/2$. Therefore, we categorize our analysis into two cases, i)  $\rho_r\neq\mu/2$, and ii) $\rho_r=\mu/2$.
\subsection{Case 1: $\rho_r\neq\mu/2$}
The transformation given by  Eq.\eqref{eq:transform} is well defined and the modified density equation corresponding to Eq.\eqref{eq:fode} in the re-scaled form reduces to
\begin{equation}
\bigg(1+\frac{1}{\sigma}\bigg)\dfrac{\partial\sigma}{\partial x}=\dfrac{\Omega\mu}{2\rho_r-\mu}.
\end{equation}
Integration of above equation yields
\begin{equation}
\label{eq:lamberteq}
\mid\sigma(x)\mid\exp\big(\sigma(x)\big)=Y(x),
\end{equation}
where $Y(x)$ is
\begin{equation}
\label{eq:y}
Y(x)=\mid\sigma(x_0)\mid\exp\Bigg(\dfrac{(x-x_0)\Omega\mu}{2\rho_r-\mu}+\sigma(x_0)\Bigg).
\end{equation}
Since the values of $\sigma(x_0)$ is known at the boundaries, we put $x_0$  equal to $0$ or $1$ which further gives
\begin{equation}
\label{eq:yalphabeta}
\begin{aligned}
Y_\alpha(x)&=\mid\sigma(0)\mid\exp\Bigg(\dfrac{x\Omega\mu}{2\rho_r-\mu}+\sigma(0)\Bigg),\\
Y_\beta(x)&=\mid\sigma(1)\mid\exp\Bigg(\dfrac{(x-1)\Omega\mu}{2\rho_r-\mu}+\sigma(1)\Bigg).
\end{aligned}
\end{equation}

 The equations in the form of Eq.\eqref{eq:lamberteq} have an explicit solution defined in terms of Lambert $W$ function \cite{corless1996lambertw} and can be written as
\begin{equation}
\label{eq:relambert}
\begin{aligned}
\sigma(x)  &=W\big(Y(x)\big),~~~~~&\sigma(x)\geq0,\\
\sigma(x)&=W\big(-Y(x)\big),~~~~~&\sigma(x)<0.
\end{aligned}
\end{equation}
The Lambert $W$ function is a multi-valued function having two real branches that are referred to as $W_0(x)$ and $W_{-1}(x)$. The branch $W_0(x)$ is defined for $x\geq -1/e$ whereas $W_{-1}(x)$ is defined in $-1/e\leq x\leq 0$  and both branches meet at $x= -1/e$.
Using the properties of the Lambert $W$ function, the solution to Eq.\eqref{eq:relambert} is obtained as
\begin{equation}
\label{eq:branch}
\sigma(x)=\begin{cases} W_{-1}(-Y) &~~~~~~~~ \sigma<-1,\\ W_0(-Y)& -1\leq\sigma<0,\\W_0(Y)&~~~~~~~~\sigma\geq0.
\end{cases}
\end{equation}
The solution thus obtained is in terms of the re-scaled density $\sigma(x)$ which can be transformed back to obtain $\rho(x)$. Analogous to the open TASEP with LK \cite{parmeggiani2004totally}, $\rho_\alpha(x)$  is stable only for $\alpha_{eff}\leq1/2$ and is always in low density regime i.e., $\rho_\alpha(x)\leq1/2$. Similarly, $\rho_\beta(x)$ is stable only when $\beta_{eff}\leq1/2$ and displays high density regime i.e, $\rho_\beta(x)\geq1/2$. In the following result, we obtain the suitable branch in terms of Lambert-W function.
\\
\textbf{\textit{Result 1:}}
For given $\Omega$, $\mu$, $\alpha$ and $\beta$, $\rho_\alpha(x)$ and $\rho_\beta(x)$ are determined as follows:\\
\label{result2}
(i) If $\rho_r>\frac{\mu}{2}$ holds, then
\begin{equation}
\label{eq:resgreaterthanhalf}
\begin{aligned}
\rho_\alpha(x) &=\frac{1}{2}\Bigg[W_{-1}\big(-Y_\alpha\big)\Bigg(\frac{2\rho_r}{\mu}-1\Bigg)+\frac{2\rho_r}{\mu}\Bigg],\\
\rho_\beta(x) &=\frac{1}{2}\Bigg[W_{0}\big(Y_\beta\big)\Bigg(\frac{2\rho_r}{\mu}-1\Bigg)+\frac{2\rho_r}{\mu}\Bigg].
\end{aligned}
\end{equation}
(ii) If $\rho_r<\frac{\mu}{2}$, then
\begin{equation}
\label{eq:reslesserthanhalf}
\begin{aligned}
	\rho_\alpha(x) &=\frac{1}{2}\Bigg[W_{0}\big(Y_\alpha\big)\Bigg(\frac{2\rho_r}{\mu}-1\Bigg)+\frac{2\rho_r}{\mu}\Bigg],\\
	\rho_\beta(x) &=\frac{1}{2}\Bigg[W_{-1}\big(-Y_\beta\big)\Bigg(\frac{2\rho_r}{\mu}-1\Bigg)+\frac{2\rho_r}{\mu}\Bigg].
\end{aligned}
\end{equation}
\\

\textit{Proof:} Let $\rho_r>\frac{\mu}{2}$. Since $\rho_\alpha(x)$ is bounded above by $1/2$, utilizing the transform from Eq.\eqref{eq:transform}, we obtain  $\sigma_\alpha(x)\leq-1$. According to Eq.\eqref{eq:branch}, the re-scaled  density corresponding to the $\rho_\alpha(x)$ is given by
\begin{equation}
\sigma_\alpha(x)=W_{-1}\big(-Y_\alpha\big)
\end{equation}
Similarly, by employing Eq.\eqref{eq:transform},  $\rho_\beta(x)\geq 1/2$ transforms to  $\sigma_\beta(x)\geq-1$. Comparing with  Eq.\eqref{eq:branch}  yields the following:
\begin{equation}
\begin{aligned}
\sigma_\beta(x)=\begin{cases} W_0\big(-Y_\beta\big) & \sigma_\beta(x)\in[-1,0),\\W_0\big(Y_\beta\big)&  \sigma_\beta(x)\geq 0.
\end{cases}
\end{aligned}
\end{equation}
We claim that $\sigma_\beta(x)=W_0\big(Y_\beta\big)$. Let us assume, on contrary, that $\sigma_\beta(x)\in[-1,0)$. Simple calculations reveal that $\rho_\beta(x)<\rho_r$ for all $x\in[0,1]$. In particular, it is known that $\rho_\beta(1)=1-\beta_{eff}$ which implies $\mu<\rho_r$. This is clearly impossible which prompts that our assumption is wrong and $\sigma_\beta(x)\notin[-1,0)$. Hence,
\begin{equation}
\sigma_\beta(x)=W_0\big(Y_\beta\big).
\end{equation}
The above expressions for $\sigma_\alpha(x)$ and $\sigma_\beta(x)$ together with Eq.\eqref{eq:transform} leads to Eq.\eqref{eq:resgreaterthanhalf}.
An analysis on similar lines yields Eq.\eqref{eq:reslesserthanhalf} and concludes the proof.
\vspace{0.5cm}

Further, the density profiles are procured depending upon how both the solutions are matched. For this purpose, we utilize the continuity of current \cite{parmeggiani2004totally} and obtain a point $x_w$ at which the density profile becomes discontinuous. This yields $\rho_{\alpha}(x_w)=1-\rho_{\beta}(x_w)$  and the density profile is then expressed as follows:
\begin{enumerate}
\item When $x_w\leq0$, the density is expressed purely in terms of $\rho_\alpha(x)$.
\item If $x_w\in(0,1)$, the density is prescribed by combination of $\rho_\alpha(x)$ and $\rho_\beta(x)$, and is obtained as
\begin{equation}
\begin{aligned}
\rho(x)=&\begin{cases}  \rho_\alpha(x)&  x\leq x_w,\\ \rho_\beta(x) & x> x_w.
\end{cases}
\end{aligned}
\end{equation}
\item For $x_w\geq 1$, the entire density profile is given by $\rho_\beta(x)$.
\end{enumerate}
Here, $\rho_{\alpha}(x)$ and $\rho_{\beta}(x)$ are determined by Eq.\eqref{eq:resgreaterthanhalf} and \eqref{eq:reslesserthanhalf}.
\subsection{\label{try}Case 2: $\rho_r=\mu/2$}
Now, for a fixed $\mu$ and $\Omega$, we intend to procure the relation between $\alpha$ and $\beta$ corresponding to which the reservoir density is half the filling factor. For this purpose, we set $\rho_r=\mu/2$ in Eq.\eqref{eq:reseq} and obtain
\begin{equation}
\label{eq:alphabetarel}
(\beta-\alpha)(2-\beta-\alpha)=2\Omega(1-\mu).
\end{equation}
If $\alpha$ and $\beta$ satisfy the above equation, the transform given by Eq.\eqref{eq:transform} is not defined. However, in such a situation, Eq.\eqref{eq:fode} simplifies to
\begin{equation}
(2\rho-1)\bigg(\dfrac{\partial \rho}{\partial x}-\dfrac{\Omega}{2}\bigg)=0.
\end{equation}
The above equation possesses two distinct solutions: a constant density $\rho_{\text{MC}}(x)=1/2$ corresponding in maximal current phase of TASEP and a linear profile $\rho(x)=\frac{\Omega x}{2}+C$.
Incorporating the boundary conditions prescribed in Eq.\eqref{eq:bc}, we can have two possible linear profiles given as
\begin{equation}
\begin{aligned}
\rho_\alpha(x)&=\frac{\Omega x+\alpha}{2}\\
\rho_\beta(x)&=\frac{\Omega(x-1)+2-\beta}{2}
\end{aligned}
\end{equation}
It is evident that attachment and detachment rates are equal in this case and is given by $\Omega/2$. Our expressions agree with that obtained in  the open TASEP with LK and constant reservoir  \cite{parmeggiani2004totally}.
Clearly, the boundary conditions impose the density less than $1/2$ at $x=0$ and greater than $1/2$ at $x=1$ for $0\leq\alpha,\beta\leq 1$. Depending on how $\rho_\alpha(x)$, $\rho_{\text{MC}}(x)$ and $\rho_\beta(x)$ are matched, different  scenarios for the density profiles appear \cite{parmeggiani2004totally}. The linear profile $\rho_\alpha(x)$ is separated from $\rho_{\text{MC}}(x)$ at $x_\alpha=(1-\alpha)/\Omega\geq 0$; whereas $\rho_{\text{MC}}(x)$ and $\rho_\beta(x)$ is distinguished at $x_\beta=(\beta+\Omega-1)/\Omega\leq 1$. Depending upon the values of $x_\alpha$ and $x_\beta$, the density profiles can be obtained as follows:

\begin{enumerate}
\item When $x_\alpha\leq x_\beta$, the density profile is expressed as
\begin{equation}
\begin{aligned}
\rho(x)=\begin{cases} \frac{\Omega x+\alpha}{2} & 0\leq x\leq x_\alpha,\\ \frac{1}{2}&  x_\alpha\leq x\leq x_\beta,\\ \frac{\Omega(x-1)+2-\beta}{2} & x_\beta\leq x\leq1.
\end{cases}
\end{aligned}
\end{equation}
The density profile is continuous and exhibits coexistence of all three phases.
\item For  $x_\alpha> x_\beta$, the density profile admits a discontinuity at a point $x_w$, where the current corresponding to $\rho_\alpha(x)$ and $\rho_\beta(x)$  match. If $x_w\leq0$ $(x_w\geq1)$, then the density profile is described by $\rho_\beta(x)$ ($\rho_\alpha(x)$), whereas  $x_w\in(0,1)$ leads to a shock in the bulk. The density profile is given by
\begin{equation}
\begin{aligned}
\rho(x)=\begin{cases} \frac{\Omega x+\alpha}{2} & 0\leq x\leq x_w,\\ \frac{\Omega(x-1)+2-\beta}{2} & x_w\leq x\leq1.
\end{cases}
\end{aligned}
\end{equation}
\end{enumerate}

For any the possible values of reservoir density, $\rho_\alpha(x)$ and $\rho_\beta(x)$ correspond to low density and high density branch, respectively.
Before concluding this section, an important feature of our model, that holds for any $\Omega$ and $\mu$ irrespective of the cases discussed above, is worth mentioning.\\
\textbf{\textit{Result 2:}}
\label{result1}
Langmuir isotherm is always achieved at $\alpha=1$ and $\beta=1$ for any value of $\mu$ and $\Omega$.\\
\textit{Proof.} At $\alpha=1$ and $\beta=1$, the densities at both the boundaries are equal and is  given by
\begin{equation}
    \rho(0)=\rho(1)=\frac{\rho_r}{\mu}.
\end{equation}

When  $\rho_r\neq\mu/2$, utilizing the above values of $\rho(0)$ and  $\rho(1)$ in Eq.\eqref{eq:yalphabeta} yields
\begin{equation}
   Y_\alpha(x)=0 \text{ and } Y_\beta(x)=0
\end{equation}
which, in turn, implies that $\sigma_\alpha(x)=\sigma_\beta(x)=0$. Thus, from Eq.\eqref{eq:transform}, we obtain an integral equation whose solution is
\begin{equation}
    \rho(x)=\frac{\mu}{1+\mu}.
\end{equation}
The above density represents Langmuir isotherm.\\
For $\rho_r=\mu/2$, the boundary conditions are reduced to $\rho(0)=\rho(1)=1/2$. Also, it can be readily calculated that $x_\alpha=0$ and $x_\beta=1$. Thus, the entire density profile is expressed in terms of $\rho_{\text{MC}}(x)$. From Eq.\eqref{eq:onerhor}, it is revealed that $\mu=1$, and thus,  $\rho_{\text{MC}}(x)$ corresponds to Langmuir isotherm. This concludes the proof.
\vspace{0.5cm}

In the following section, we obtain the theoretical existence conditions of different phases in the steady state.
\section{\label{sec:level4}Existence of stationary phases}

We first discuss the possible phases that can arise in the density profiles. When $\rho_r\neq\mu/2$, clearly the effective rate of attachment and detachment are distinct in the steady-state. From Ref.\cite{evans2003shock}, we deduce that the possible phases are low density ($\text{LD}$), high density ($\text{HD}$) and Shock ($\text{S}$) phase. For $\rho_r=\mu/2$, the effective attachment and detachment rates are equal in the steady state which implies that the possible phases are $\text{LD}$, $\text{HD}$, $\text{MC}$, $\text{S}$ and $\text{LD-MC-HD}$ \cite{evans2003shock,parmeggiani2004totally}.
We now derive the condition of existence of above discussed phases.
\begin{enumerate}
\item \textbf{LD phase:} In low density regime, $\rho(x)\leq 1/2$ and $\int_0^1\rho(x)dx<1/2$. Thus, the entire density profile is expressed in terms of $\rho_\alpha(x)$. This phase exists when $\alpha$ and $\beta$ satisfies
\begin{equation}
    \int_0^1\rho_\alpha(x)dx=\mu-\rho_r.
\end{equation}
\item \textbf{MC phase:} In this phase, the lattice is exactly half filled i.e., $\int_0^1\rho(x)=1/2$. For the existence of maximal current, both sides of Eq.\eqref{eq:fode} must be simultaneously equal to $0$ \cite{evans2003shock}, resulting in $\rho=1/2$ and $\rho=\mu/(1+\mu)$. Clearly, this is possible only when $\mu=1$. 

\item \textbf{HD phase:} In high density regime,  $\rho(x)\geq 1/2$ and $\int_0^1\rho(x)dx>1/2$. the density profile is obtained by only $\rho_\beta(x)$ and its existence is ensured when the following holds for $\alpha$ and $\beta$
\begin{equation}
    \int_0^1\rho_\beta(x)dx=\mu-\rho_r.
\end{equation}
\item \textbf{S phase:} When lattice is in Shock phase, the density profile displays a discontinuity at $x_w\in (0,1)$ wherein the density at left of $x_w$ is expressed by $\rho_\alpha(x)$ and at right of $x_w$ is given by $\rho_\beta(x)$ i.e., segment to the left and right show LD and HD phase, respectively. This phase occurs when following holds
\begin{equation}
    \int_0^{x_w}\rho_\alpha(x)dx+\int_{x_w}^1\rho_\beta(x)dx=\mu-\rho_r
\end{equation}
\item \textbf{LD-MC-HD phase:} For the existence of this phase, it is necessary that the reservoir density is exactly equal to $\mu/2$. Thus, $\alpha$ and $\beta$ must satisfy Eq.\eqref{eq:alphabetarel}. Additionally, as discussed in previous section, $x_\alpha\leq x_\beta$ leads to coexistence of the three phases. Thus, the condition of existence is given by
\begin{equation}
\label{eq:ldmchd}
    \alpha+\beta+\Omega\geq 2
\end{equation}
where $\alpha$ and $\beta$ are consistent with  Eq.\eqref{eq:alphabetarel}.
\end{enumerate}
In the following section, we investigate the influence of the additional dynamics on the phase diagrams. Precisely, we scrutinize how $\mu$ and $\Omega$ impact the phase diagrams.

\begin{figure*}[!htb]
\centering
\subfigure[\label{mulesserthan1}]{\includegraphics[width = 0.4\textwidth]{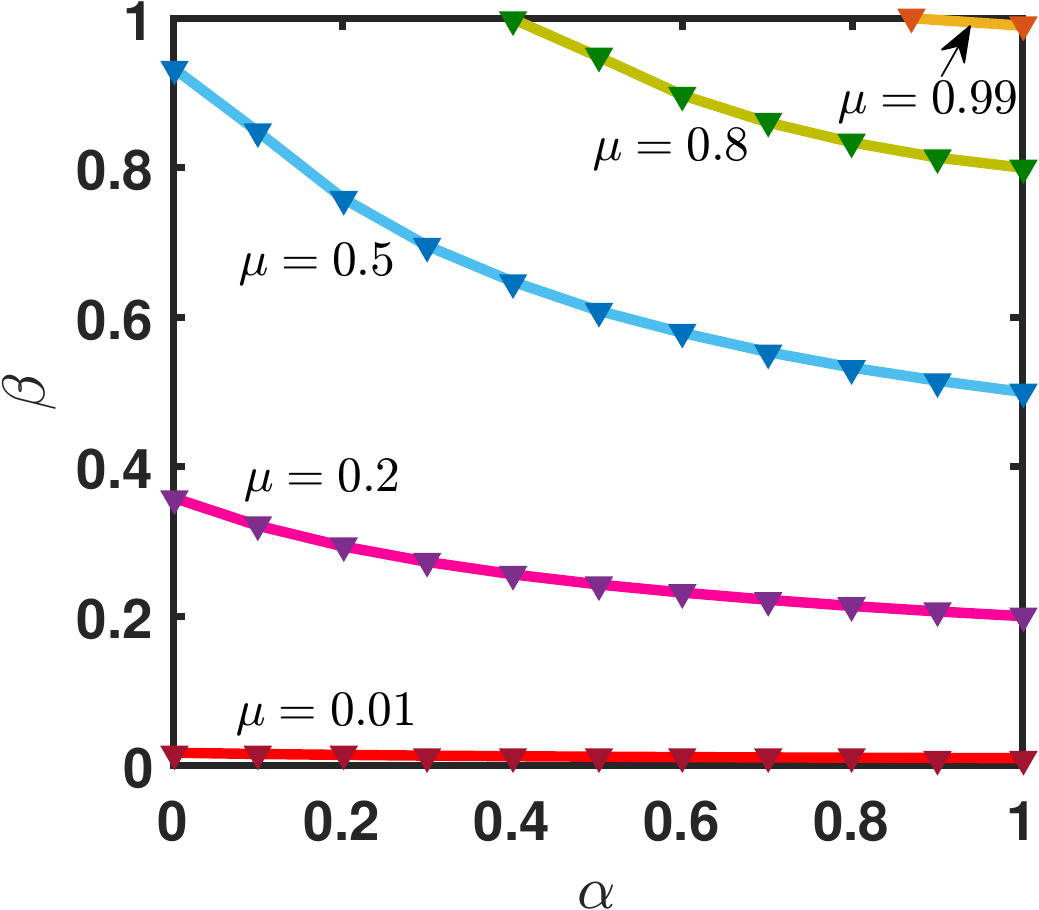}}
\subfigure[\label{mu_p5_omega_varies}]{\includegraphics[width = 0.4\textwidth]{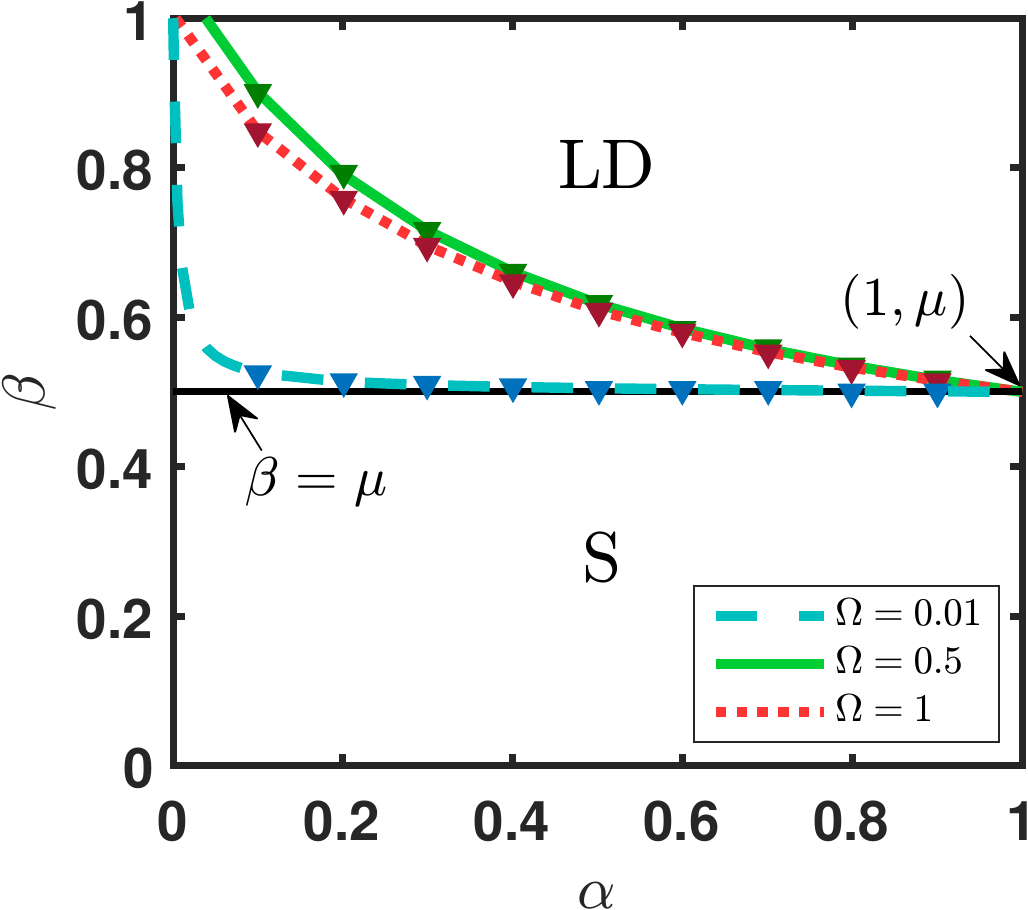}}

\caption{\label{mulessthan1}Phase diagrams for different values of $\mu$ and $\Omega$.  The region above phase boundary is $\text{LD}$ whereas $\text{S}$ exists in the region below the phase boundary.  Lines of different styles and symbols denote Mean-field and MCS results, respectively. (a) $\Omega=1$ whereas $\mu$ varies, (b) $\Omega$ is varied while $\mu=0.5$.  Black solid line denotes the phase boundary between $\text{LD}$ and  $\text{S}$ when $\Omega=0$.}
\end{figure*}
\begin{figure*}[!htb]
\centering
\subfigure[\label{shockmovement_mul1_2_new}]{\includegraphics[width = 0.4\textwidth]{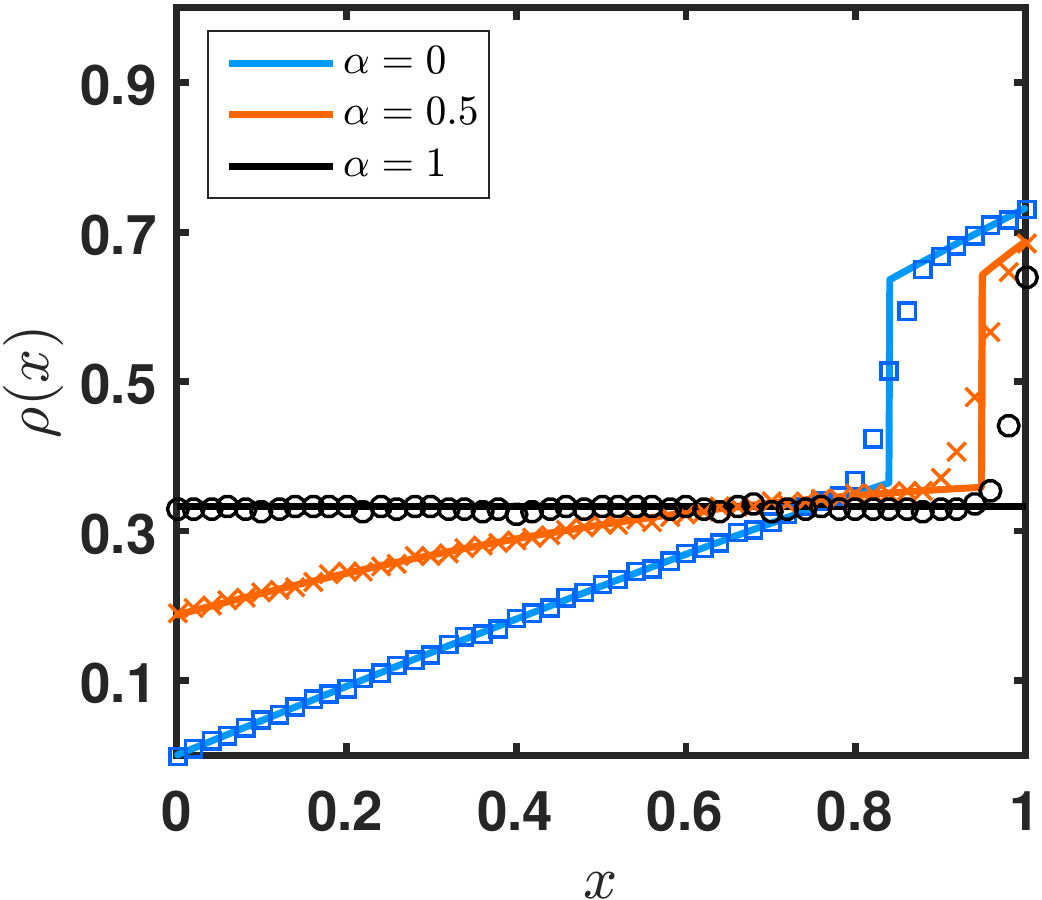}}
\subfigure[\label{shockpositionmul1_new}]{\includegraphics[width = 0.4\textwidth]{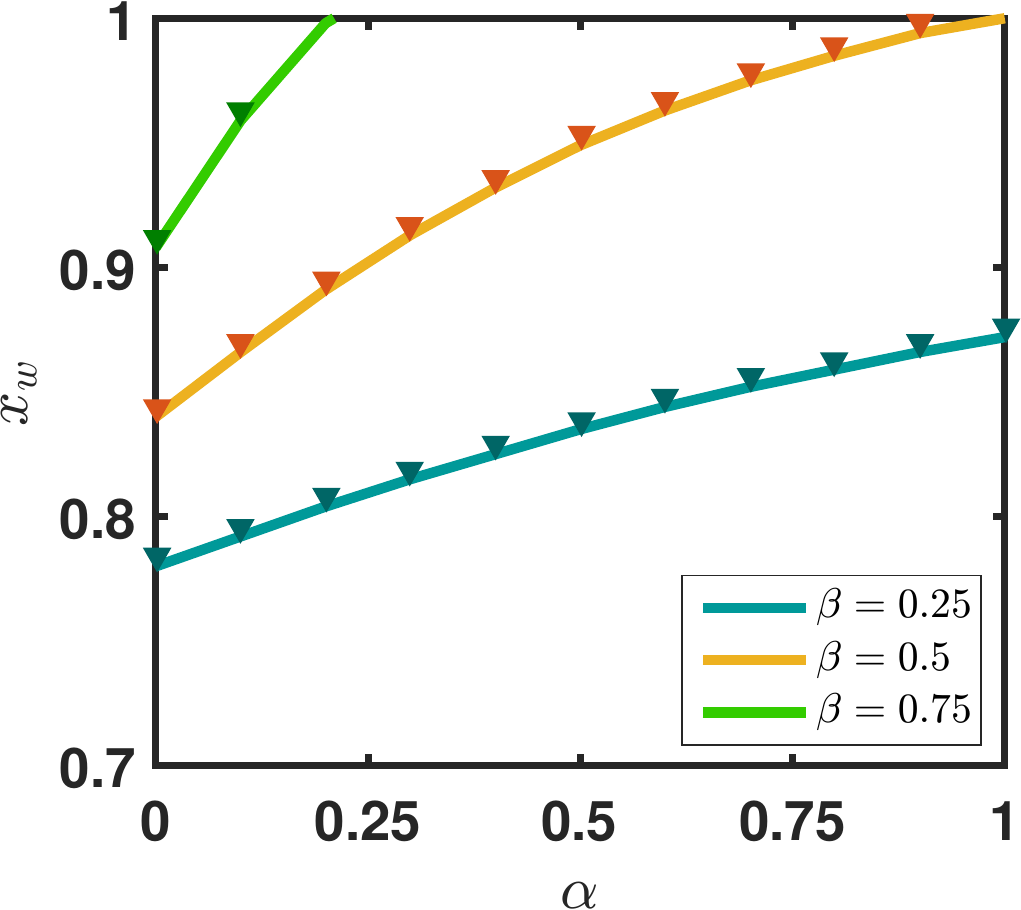}}

\caption{\label{mup5shock}(a) Density profiles demonstrating the movement of shock towards right when $\alpha$ is varied. Other parameters are $\mu=0.5$, $\Omega=1$ $\beta=0.5$, (b)  Variation of shock position, $x_w$, when $\alpha$ is varied where $\mu=0.5$ and $\Omega=1$. Solid lines and symbols denote Mean-field and MCS results, respectively.}

\end{figure*}
\section{Effect of $\mu$ and $\omega$ on phase diagram}
To study the effect of the total number of particles and LK on the properties of system in steady state, we derive phase diagrams theoretically as discussed in previous section for specific values of $\mu$ and $\Omega$ in the parameter space of $\alpha$-$\beta$. To validate our theoretical findings, we use Monte Carlo Simulations (MCS)  with random sequential update rule.
Here, a site is randomly selected and updated in accordance with the dynamical rules as defined in section \ref{sec:level1}. The lattice length is taken to be $L = 500$ and the simulations are run for $10^8$ time-steps. To facilitate the onset of a steady state, we ignore the first $5\%$ of time steps and the average particle density is calculated for an interval of $10L$.  The filling factor $\mu$ represents the average number of particles available for each lattice site, and therefore, it is expected that $\mu$ will significantly affect the phase diagram. So, we categorize into three subsections, namely, $\mu<1$, $\mu=1$ and $\mu>1$.
\subsection{$\mu<1$}
We obtain the phase diagram for different values of $\mu<1$ and $\Omega$ in the parameter space of $\alpha$ and $\beta$ (see Fig.\ref{mulessthan1}). The phase diagram for any choice of $\mu<1$ and $\Omega$ contains only two distinct phases,  $\text{LD}$ and $\text{S}$.
The boundary between these phases can be obtained by utilizing the current continuity principle which yields $\rho_\alpha(1)\big(1-\rho_\alpha(1)\big)=\rho_\beta(1)\big(1-\rho_\beta(1)\big)$. Since $\rho_\beta(1)=1-\beta_{eff}$, we obtain
\begin{equation}
\label{eq:Jcontmul1}
\rho_\alpha(1)\big(1-\rho_\alpha(1)\big)=\beta\bigg(1-\dfrac{\rho_r}{\mu}\bigg)\Bigg(1-\beta\bigg(1-\dfrac{\rho_r}{\mu}\bigg)\Bigg).
\end{equation}
In the above equation, $\rho_\alpha(1)$  and $\rho_r$ depends on the controlling parameter $\alpha$ and $\beta$ leading to an implicit equation which can be solved to obtain the phase boundary in $\alpha-\beta$ plane.
\begin{figure*}[!htb]
\centering
\subfigure[\label{mueq1_new}]{\includegraphics[width = 0.4\textwidth]{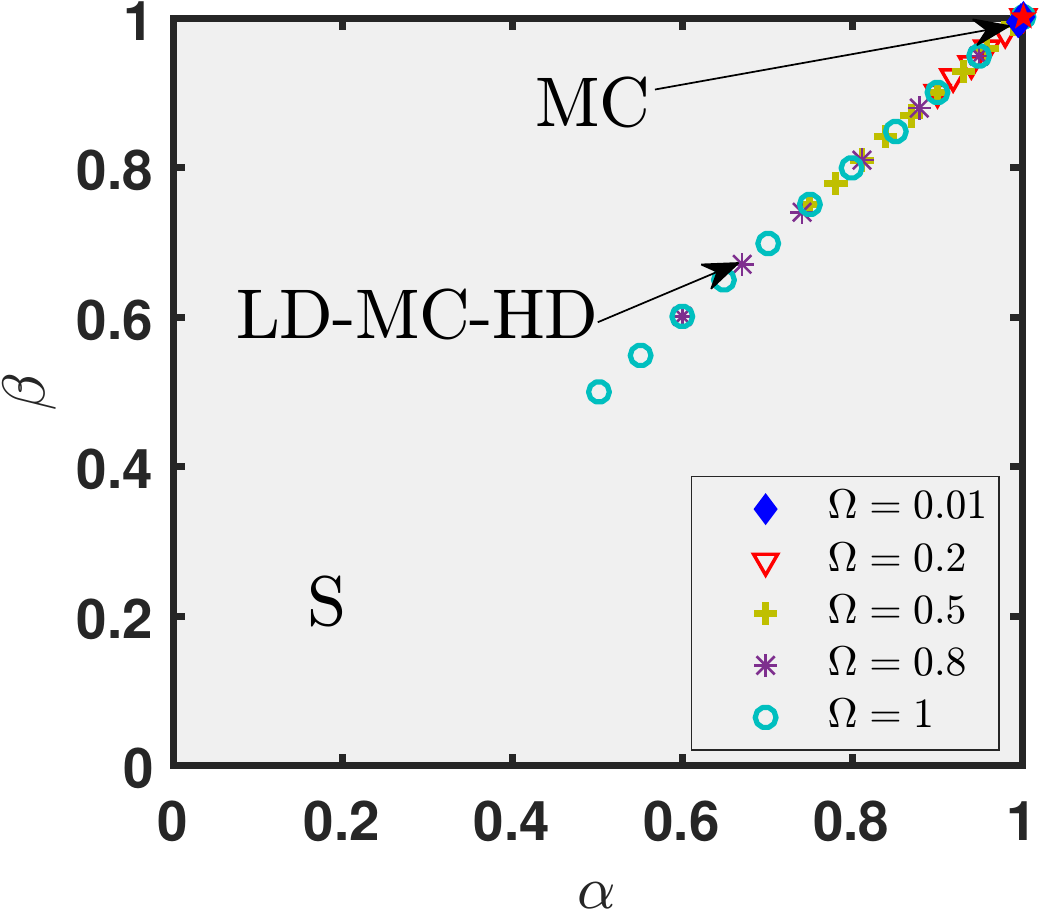}}
\subfigure[\label{mu1density}]{\includegraphics[width = 0.4\textwidth]{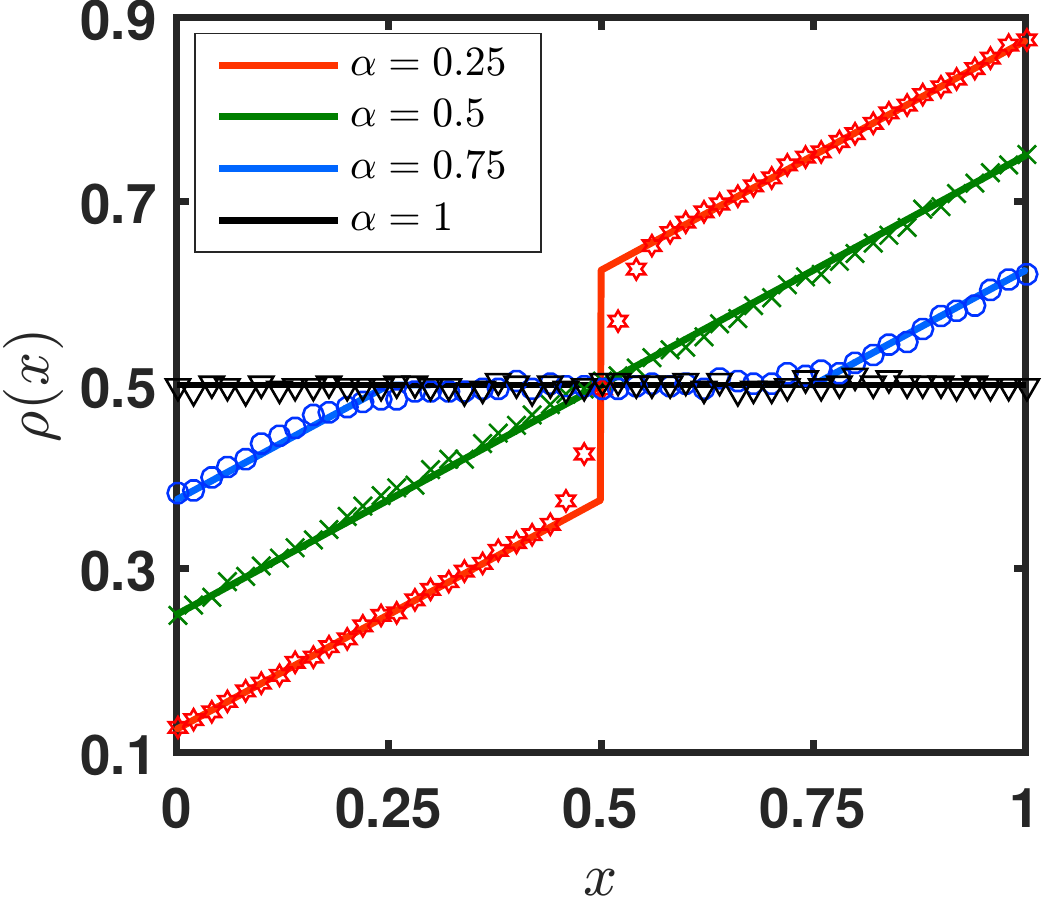}}

\caption{\label{mueq1}(a) Phase diagram for various values of $\Omega$ where $\mu=1$. Symbols denote the  $\text{LD-MC-HD}$ phase and the area shaded in gray denotes $\text{S}$ phase. $\text{MC}$ phase (denoted by red pentagram) is confined to a point $(\alpha,\beta)=(1,1)$. (b) Density profiles corresponding to points $(\alpha,\alpha)$ for  $\mu=1$, $\Omega=1$ and different $\alpha$. Solid lines and symbols denote the Mean-field and MCS results, respectively.}
\end{figure*}

Now, to explore the role of filling factor on the phase diagram, we vary $\mu<1$ while keeping $\Omega$ fixed. It is evident from Fig.\ref{mulesserthan1} that as $\mu$ increases from $0$ to $1$, the phase boundary shifts resulting in shrinkage of $\text{LD}$ phase and  expansion of $\text{S}$ phase. This observation can be explained as follows: when $\mu$ is very small, there are no particles in the reservoir leading to very low effective entry and attachment rates. As a result the phase diagram trivially exhibits $\text{LD}$ phase in the entire $\alpha-\beta$ plane. The increase in $\mu$ results in increased  effective entry and attachment rates whereas the effective exit and detachment rates are reduced due to the choice of $f(.)$ and $g(.)$. This, in turn, feeds more particles onto the lattice, and their exit is hindered. Therefore, the boundary layer enters in the bulk as a shock due to which the region containing $\text{S}$ phase widens and $\text{LD}$ region contracts.

In order to visualize the effect of LK on the phase diagram, we keep $\mu$  fixed and obtain phase diagrams for different $\Omega$.  From Fig.\ref{mu_p5_omega_varies}, it is observed that decreasing $\Omega$ from $1$ to $0$ results in a non-monotonic shift of the phase boundary. Furthermore, the phase transition line approaches $\beta=\mu$ as $\Omega\to 0$, and the model converges to that in Ref.\cite{haldar2020asymmetric}  for $0\leq \alpha,\beta\leq 1$. It is ascertained  from Fig.\ref{mu_p5_omega_varies},  that the point $(\alpha,\beta)=(1,\mu)$ always lies on a phase boundary between $\text{LD}$ and $\text{S}$ phase, for any value of $\Omega$. The following result ensures this observation.\\
\textbf{\textit{Result 3:}}
\label{result3}
For a fixed $\mu<1$, the point $(\alpha,\beta)=(1,\mu)$ lies on the phase boundary between $\text{LD}$ and $\text{S}$ phases irrespective of the values of $\Omega$.\\
\textit{Proof}: For $(\alpha,\beta)=(1,\mu)$, from Eq.\eqref{eq:reseq}, one can easily obtain two distinct values of $\rho_r$ given as
\begin{equation}
\rho_r=\frac{\mu^2}{\mu+1} \text{ and } \rho_r=\mu+\frac{\Omega\mu}{1-\mu}.
\end{equation}
Since $\rho_r$ can not exceed $\mu<1$ owing to PNC, hence we have $\rho_r=\frac{\mu^2}{\mu+1}$ which is independent of $\Omega$. Further, from Eq.\eqref{eq:bc} we compute $\rho(0)=\frac{\mu}{1+\mu}$  which subsequently yields $\sigma(0)=0$. Thus, Eq.\eqref{eq:lamberteq}  results in the following equation for $\sigma_\alpha(x)$,
\begin{equation}
\label{eq:sigmaalpha}
\mid\sigma_\alpha(x)\mid\exp\big(\sigma_\alpha(x)\big)=0.
\end{equation}
From above equation, we deduce that $\sigma_\alpha(x)=0$ leading to a constant $\text{LD}$ branch given by $\rho_\alpha(x)=\frac{\mu}{1+\mu}$. This expression gives $\rho_\alpha(1)=\frac{\mu}{1+\mu}$ that satisfies Eq.\eqref{eq:Jcontmul1}. This concludes the proof.

\begin{figure*}[!htb]
\centering
\subfigure[\label{densitymueq1}]{\includegraphics[width = 0.4\textwidth]{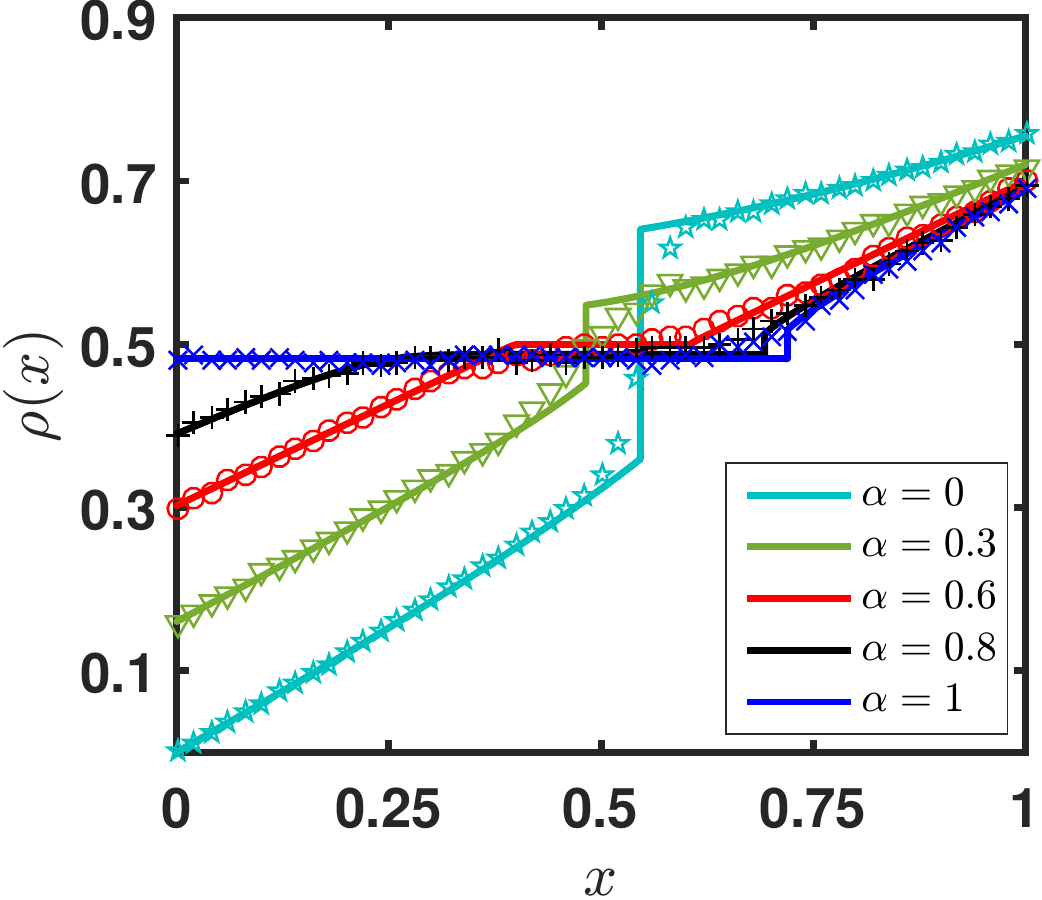}}
\subfigure[\label{mueq1transition}]{\includegraphics[width = 0.4\textwidth]{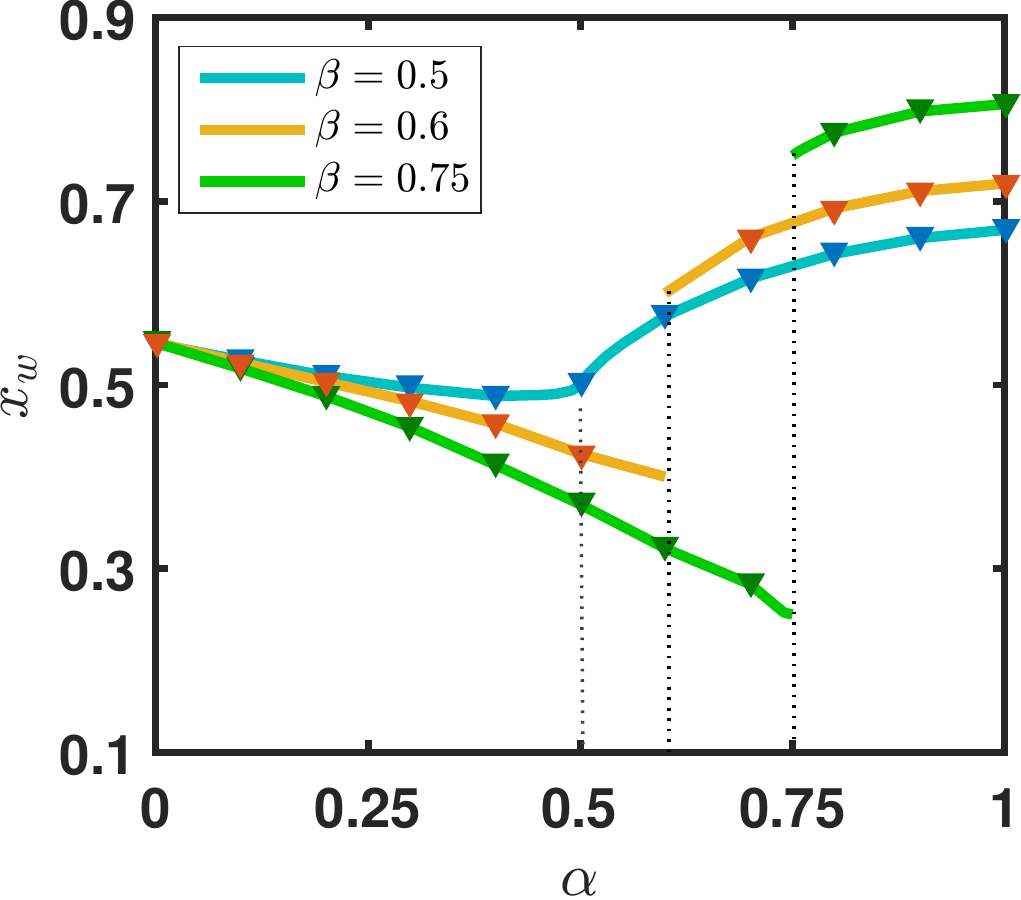}}

\caption{\label{mueq1transition1}(a) Density profiles depicting the back-and-forth phase transition for $\beta=0.6$. (b) Position of shock, $x_w$, when $\alpha$ is varied. Other parameters are $\mu=1$ and $\Omega=1$. Solid lines and symbols denote Mean-field and MCS results, respectively. Dotted lines denote the value of $\alpha$ corresponding to $\beta$ for which back-and-forth transition occurs.}
\end{figure*}
To discuss the nature of phase transitions between $\text{LD}$ and $\text{S}$  phases for a given $\mu$  and $\Omega$, we have plotted density profiles for different $\alpha$ while keeping the other parameters fixed in Fig.\ref{shockmovement_mul1_2_new}. 
With an increase in $\alpha$, $\alpha_{eff}$ increases while $\beta_{eff}$ decreases. In the absence of LK dynamics, this would have resulted in shock movement towards the left end. However, in our model, increase in $\alpha$ leads to increase in difference between $\Omega_{d_{eff}}$ and $\Omega_{a_{eff}}$ which dominates over the boundary effect causing the movement of shock towards right end. As a result of interplay between boundary dynamics and LK dynamics, the number of particles reduces on the lattice. Beyond a crucial value of $\alpha$,  shock reaches the right boundary and $\text{LD}$ phase appears. Further, we have plotted position of shock in the bulk, $x_w$, with respect to $\alpha$ in Fig.\ref{shockpositionmul1_new} for for different values of $\beta$. Here $x_w$  displays a continuous variation with respect to $\alpha$  which ensures that the phase transition from S to LD phase is of second order.

\subsection{$\mu=1$}

In this subsection, we aim to discuss the special case when number of particles in the system is equal to the total number of lattice sites i.e., $\mu=1$. We obtain the phase diagram for different $\Omega$ in the $\alpha-\beta$ parameter space. In previous subsection, it was observed that as $\mu$ approaches $1$, the S phase covers almost all the region in the phase diagram, This schema continues for $\mu=1$ and $\text{LD}$ phase completely disappears and S phase predominantly occupies the phase diagram.  Also, the non-trivial effect of $\mu=1$ emerges with the appearance of a mixed $\text{LD-MC-HD}$ and MC phases which is shown in Fig.\ref{mueq1_new}. It is visible that $\text{LD-MC-HD}$ and $\text{MC}$ phase  are restricted to a line segment contained in $(\alpha,\alpha)$. As discussed in section \ref{sec:level4}, both $\text{LD-MC-HD}$ and $\text{MC}$ phase occur only for $\rho_r=1/2$ that implies $\alpha=\beta$ from Eq.\eqref{eq:alphabetarel}. Utilizing the existence condition for $\text{LD-MC-HD}$ given in Eq. \eqref{eq:ldmchd} results in
\begin{equation}
\label{eq:lmh}
\alpha\geq1-\frac{\Omega}{2} \text{ and } \alpha=\beta.
\end{equation}
In particular for the point $(\alpha,\beta)=(1,1)$, the occurrence of $\text{MC}$ phase is ensured by Result 2. Fig.\ref{mu1density} demonstrates the variation of phases in the diagonal ($\alpha,\alpha$) in terms of density profiles with specific set of parameters and the theoretical results agree very well with MCS.


Further, the variation of $\Omega$ from $1$ to $0$ does not alter the topology of the phase diagram which remains preserved as seen in Fig.\ref{mueq1_new}. However, the line occupied by $\text{LD-MC-HD}$ phase shrinks towards $(\alpha,\beta)=(1,1)$ and in the limiting case $\Omega\to 0$, it ceases to exist. This, in turn, agrees with the fact that $\text{LD-MC-HD}$ phase is missing in the analogous model devoid of LK \cite{haldar2020asymmetric}.

Now, we discuss an important feature of back-and-forth transition that arises due to the peculiar structure of the phase diagram for $\mu=1$ \cite{brackley2012multiple,verma2019stochastic}. It has been observed that for a fixed $\beta\in[1-\Omega/2,1)$ when $\alpha$ is increased, one can pass from S  to LD-MC-HD to again S phase (S$\rightarrow$LD-MC-HD$\rightarrow$S) as also evident in Fig.\ref{mueq1_new}. We have plotted density profiles for $\beta=0.6$, $\Omega=1$  and increasing $\alpha$ in Fig.\ref{densitymueq1} which agree well with the MCS results. This behavior can be understood as follows. With an increase in $\alpha$, the difference between $\alpha_{eff}$ and $\beta_{eff}$ decreases which allows more particles on the lattice; whereas $\Omega_{a_{eff}}$  reduces and $\Omega_{d_{eff}}$ increases thereby admitting fewer particles. As a result, a competition ensues between  the boundary dynamics and LK dynamics. Thus, when  $\beta\in[1-\Omega/2,1)$ is fixed, for $\alpha<\beta$ the system is governed by the boundary dynamics that increases the number of particles on the lattice leading to the movement of shock towards the left end. At $\alpha=\beta$, $\alpha_{eff}$ and $\beta_{eff}$ become equal; likewise, $\Omega_{a_{eff}}$  and $\Omega_{d_{eff}}$  also have same value. Consequently, the shock vanishes from the system and LD-MC-HD phase emerges. When $\alpha>\beta$, LK dominates resulting in the net out-flux of particles that again  gives rise to a shock in density profile. To further visualize this feature, we have plotted the position of shock with respect to $\alpha$ for different $\beta$ in Fig.\ref{mueq1transition}. 
Analogous arguments hold for the back-and-forth transitions which arises for a fixed $\alpha\in[1-\Omega/2,1)$ and $\beta$ is varied.

\subsection{$\mu>1$}
\begin{figure*}[!htb]
\centering
\subfigure[\label{mugreaterthan1pd_new}]{\includegraphics[width = 0.4\textwidth]{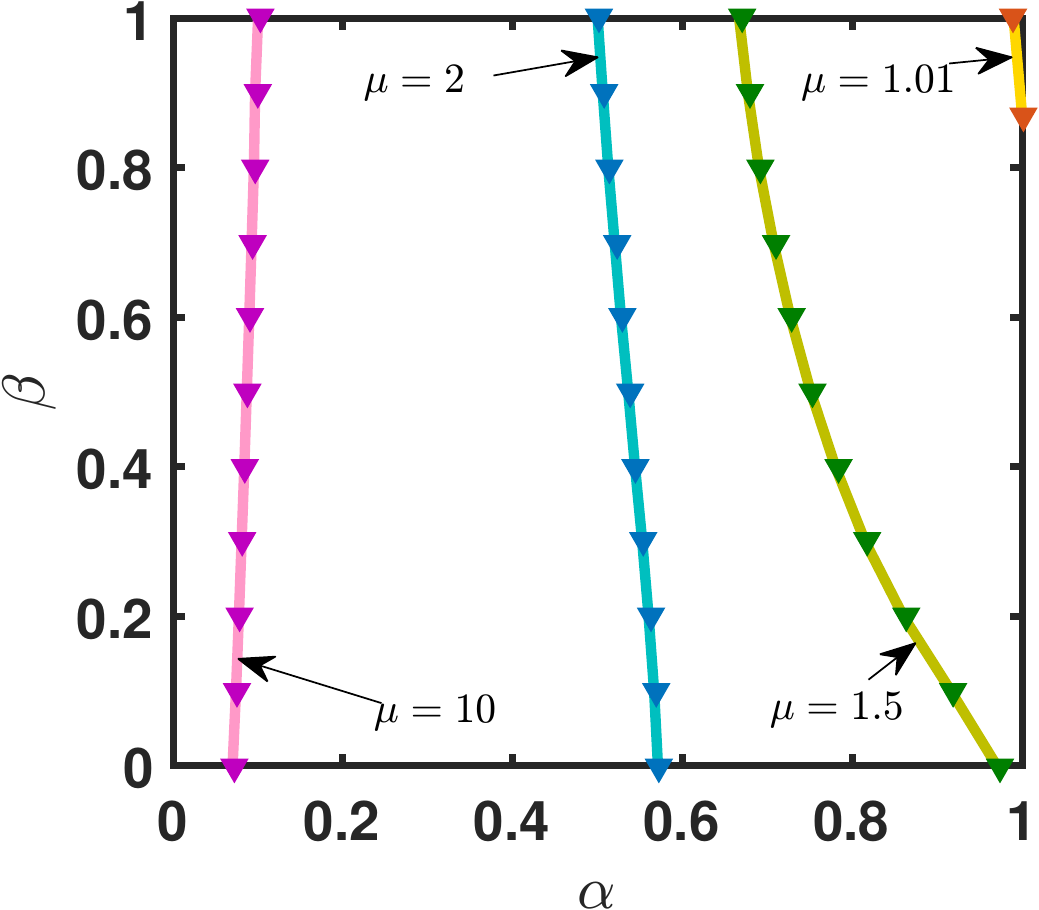}}
\subfigure[\label{mu_2_omega_varies}]{\includegraphics[width = 0.4\textwidth]{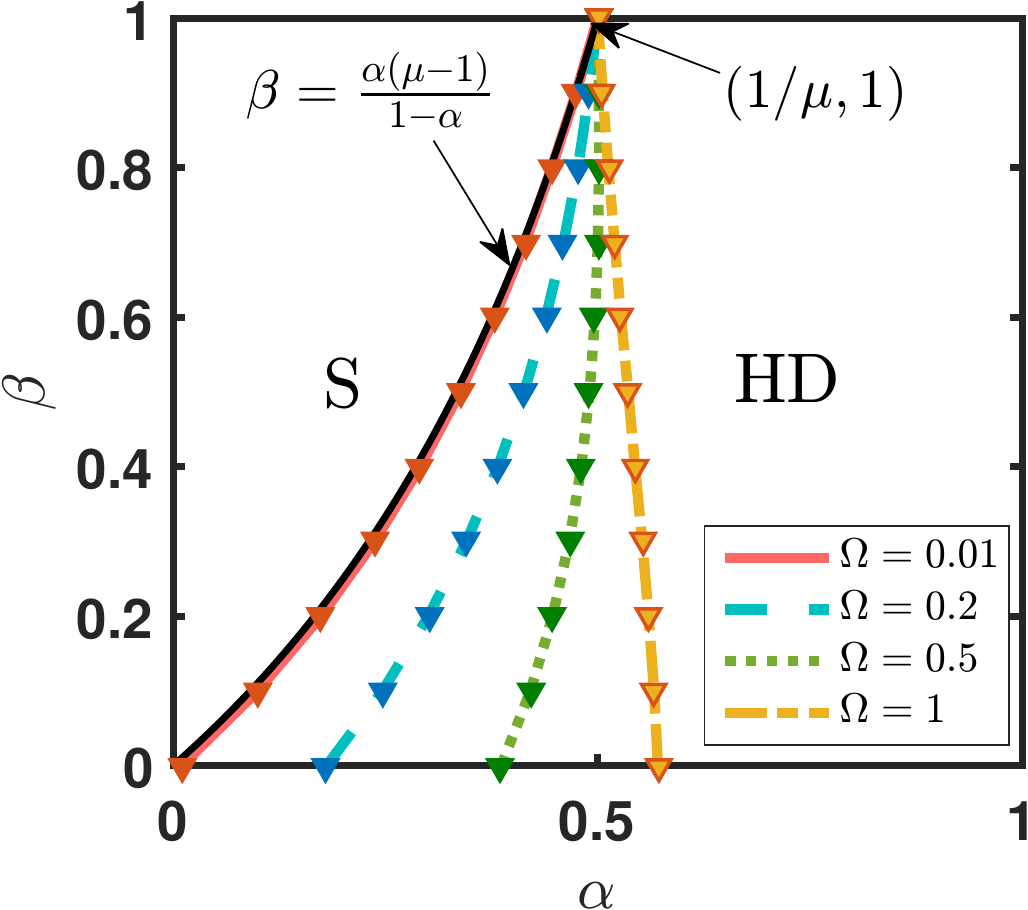}}

\caption{\label{mugreatthan1} Phase diagrams for various values of $\mu$ and $\Omega$. Lines of different styles and symbols  denote phase boundary obtained by Mean-field and MCS, respectively.  $\text{HD}$ phase lies on the right region and $\text{S}$ exists  on the left region of the phase boundary. (a) $\Omega=1$ whereas $\mu$ varies, (b) $\Omega$ is varied while $\mu=2$. Black solid line denotes the phase boundary between $\text{S}$ and  $\text{HD}$ when $\Omega=0$.}
\end{figure*}

Now, when number of particles exceed the total number of sites, we observe the topological changes in the phase diagram. As opposed to the case $\mu=1$, a new HD phase appears in the phase diagram in addition to the existing S phase whereas LD-MC-HD and MC phases completely vanish which can be seen in Fig.\ref{mugreatthan1}.  
Similar to the reasoning in $\mu<1$, the boundary between S and HD phases can be given by
 \begin{equation}
\rho_\beta(0)\big(1-\rho_\beta(0)\big)=\dfrac{\alpha\rho_r}{\mu}\bigg(1-\dfrac{\alpha\rho_r}{\mu}\bigg).
\end{equation}

In order to understand how filling factor affects the system, we alter $\mu>1$ while keeping $\Omega$ fixed. One can observe from Fig.\ref{mugreaterthan1pd_new} that as $\mu$ approaches $1$, the phase diagram predominantly displays $\text{S}$ phase. On increasing $\mu$,  the phase boundary is re-positioned due to contraction and expansion of $\text{S}$ and $\text{HD}$ phases, respectively. This can be physically attributed to the increased effective rates of entry and attachment with a simultaneous reduction in exit and detachment rates which favors the $\text{HD}$ phase.
As a consequence, in the limiting case $\mu\to\infty$ due to abundance of particles, $\text{HD}$ phase dominates entire phase diagram.

With an intend to investigate the effect of LK on the system, we construct the phase diagrams for different $\Omega$ while keeping $\mu>1$  fixed. On decreasing the value $\Omega$,  no topological change is observed in the phase diagram as indicated in Fig.\ref{mu_2_omega_varies} for $\mu=2$. Also, the boundary between $\text{S}$ and $\text{HD}$ phase approaches the curve $\beta=\alpha(\mu-1)/(1-\alpha)$ as $\Omega\to0$ and the model converges to that discussed in Ref.\cite{haldar2020asymmetric} for $0\leq\alpha,\beta\leq1$. Moreover from Fig.\ref{mu_2_omega_varies}, it is visible that the point $(\alpha,\beta)=(1/\mu,1)$ lies on the phase boundary between $\text{S}$ and $\text{HD}$ phase, for any value of $\Omega$. This observation is stated as a result below.\\
\textbf{\textit{Result 4:}}
For a fixed $\mu>1$,  the point ($\alpha,\beta)=(1/\mu,1)$  lies on the phase boundary regardless of the value of $\Omega$.\\
\textit{Proof}: Arguments similar to Result 3 holds.
\newline

\begin{figure*}[!htb]
\centering
\subfigure[\label{shockmovement_mug1_2_new}]{\includegraphics[width = 0.4\textwidth]{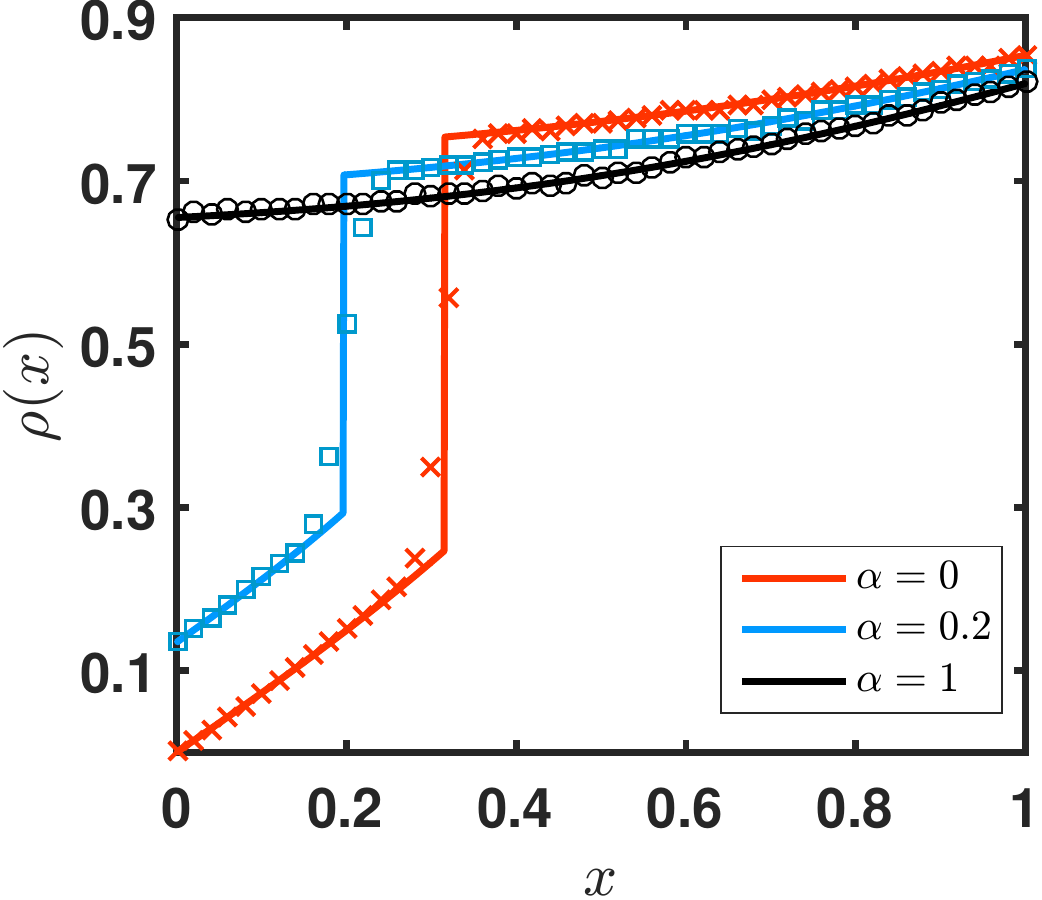}}
\subfigure[\label{shockpositionmug1_new2}]{\includegraphics[width = 0.4\textwidth]{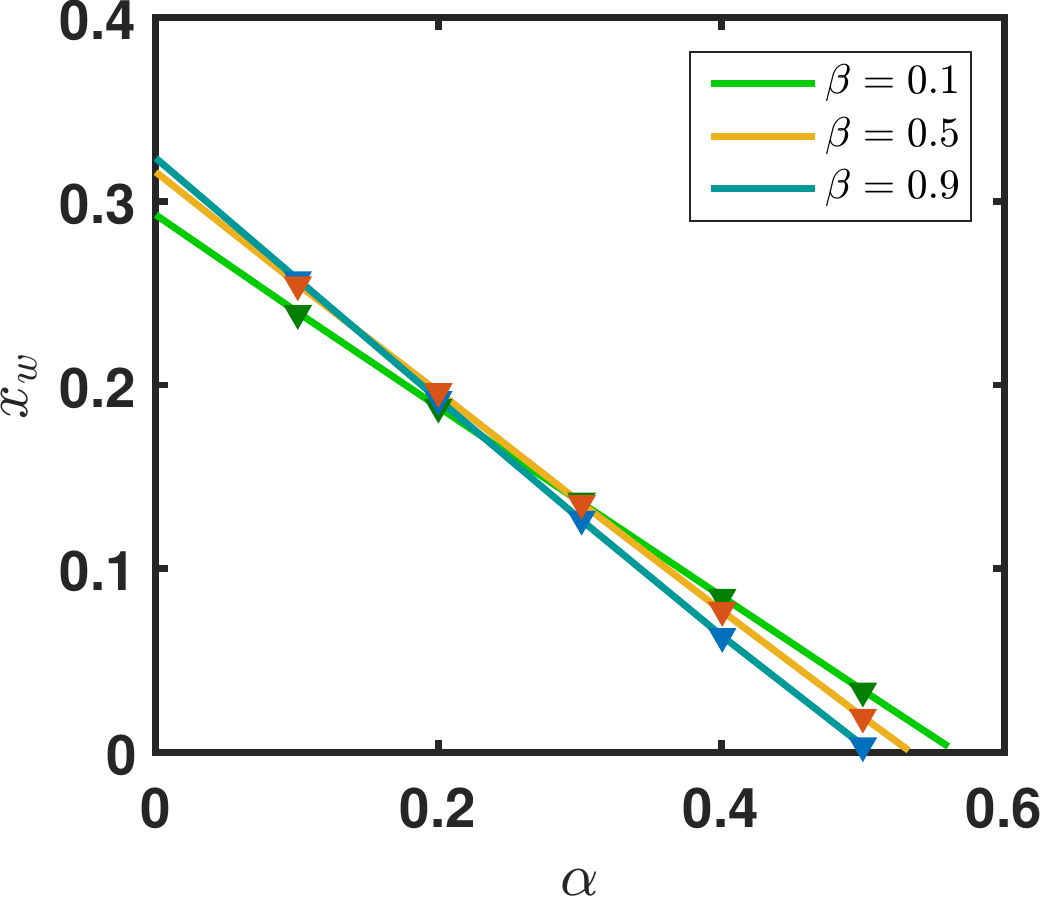}}

\caption{\label{mu2shock}  Density profiles demonstrating the movement of shock towards left when $\alpha$ is varied. Other parameters are $\mu=2$, $\Omega=1$ and $\beta=0.5$. (b) Variation of shock position ($x_w$) with respect to $\alpha$. Symbols denote MCS results and solid lines represent Mean-field results.}
\end{figure*}

We now discuss the order of phase transition between $\text{S}$ and $\text{HD}$ phases for a given $\mu$ and $\Omega$. As seen from Fig.\ref{shockmovement_mug1_2_new}, for $\beta=0.5$, an increase in $\alpha$ results in the movement of shock towards left end which reaches the left boundary at a critical value of $\alpha$. This is because $\alpha_{eff}$ and $\Omega_{d_{eff}}$  increase whereas $\beta_{eff}$ and $\Omega_{a_{eff}}$ decrease. In this interplay, boundary dynamics dominate over LK dynamics, accommodating more number of particles onto the lattice. Thus, the shock moves towards left and the segment exhibiting HD phase increases. Furthermore, the transition through the phase boundary is second order with respect to shock position which is shown in  Fig.\ref{shockpositionmug1_new2}.

To summarize, we found that the phase diagram differs qualitatively in the neighborhood of $\mu=1$ and displays the following phases: i) $\mu<1$: LD and S phases, ii) $\mu=1$: S, LD-MC-HD and MC phases, and iii) $\mu>1$: S and HD phases.  In addition to this, it was found that variation of $\Omega$ has only quantitative effect on the structure of phase diagram corresponding to the specific value of $\mu$. Interestingly, in all three cases the topology of the phase diagram is quite simple and also the number of phases remains independent of $\Omega>0$ when compared to the open TASEP with LK subjected to equal innate attachment-detachment rates \cite{parmeggiani2004totally,evans2003shock}. Furthermore, we have analyzed the transition of phases across the boundaries which turned out to be second order for $\mu\neq1$, whereas a  back-and-forth transition emerges for $\mu=1$. All the findings have been aided with intuitive explanations together with the analysis through theoretically computed phase boundaries and validated via extensive Monte Carlo simulations.

\section{Conclusion}
In this work, we have studied a variant of closed TASEP consisting of a lattice whose both ends are connected to a reservoir. Particles are allowed to attach in the bulk  from the reservoir, and can leave  the lattice to  rejoin the reservoir. Due to the closed nature of the system, the total number of particles remains conserved which is characterized by filling factor. The rates at which the particles enter and exit the lattice are regulated by the occupancy of the reservoir. Our model majorly differs from previous studies involving LK which is attributed to the consideration of reservoir crowding in the conserved system.

We investigated the steady-state properties in the theoretical framework of mean-field approximation. To reduce the parameter space and make the calculations analytically solvable, we considered the innate attachment and detachment rates to be equal. Subsequently, we derived the condition of existence of various phases and obtained the phase diagrams. It is observed that the phase diagram has two distinct phases for $\mu\neq 1$ and three different phases for $\mu=1$. 
 While considering an additional feature of reservoir crowding in an open TASEP with LK, one might naively expect to obtain a more complex phase diagram. On contrary, it is found to have a relatively simple structure with reduced number of phases. This simplification occurs due to the non-trivial effect of the reservoir crowding which serves as an intrinsic control on all the rates except the forward hopping rate. Additionally, to study the effect of total number of particles in the system, we observed the variation of the phase diagram with respect to $\mu$ for a fixed $\Omega$. It turns out that the topology of the phase diagram remains preserved except in the neighborhood of $\mu=1$. Furthermore, it is noted that varying the innate attachment / detachment rate with fixed filling factor does not alter the intrinsic features of phase diagram.
 To obtain insight into the nature of transitions across the phase boundaries, we have taken position of shock to be the order parameter. For $\mu\neq  1$, the transition across the phase boundary is found to be of second order. A significant finding of our study is the existence of back-and-forth transition which occurs only when $\mu=1$. In this case, we witnessed transitions from S to LD-MC-HD phase, and then back to S phase.

 To confirm the theoretical findings, we simulated our proposed model using Monte Carlo Simulations following a random sequential update rule. The present work is an attempt to understand the interplay between LK and filling factor in presence of the reservoir crowding and highlights some of their non trivial effects on the system dynamics.

%

\end{document}